\documentclass[aps,prl,twocolumn,amsmath,amssymb,nofootinbib,superscriptaddress]{revtex4}
\usepackage{graphicx}
\usepackage{epstopdf}
\newcommand{\bra}[1]{\langle#1|}
\newcommand{\ket}[1]{|#1\rangle}
\newcommand{\eqn}[1]{Eq.~(\ref{#1})}
\newcommand{\fig}[1]{Fig.~\ref{#1}}
\newcommand{\tpw}{T_\mathrm{TPW}}
\newcommand{\tr}{\mathrm{Tr}}

\begin{document}

\title{Quantum Limits of Thermometry}

\author{Thomas M. Stace} \email[]{stace@physics.uq.edu.au}
\affiliation{Department of Physics, University of Queensland, Brisbane, QLD 4072, Australia}
\date{\today}
\frenchspacing

\begin{abstract}
The precision of typical thermometers consisting of $N$ particles is shot noise limited, improving as $\sim1/\sqrt{N}$.  For high precision thermometry and thermometric standards this presents an important theoretical noise floor. 
 Here it is demonstrated that thermometry may be mapped onto the problem of phase estimation, and using techniques from optimal phase estimation, it follows that the scaling of the precision of a thermometer may in principle be improved to $\sim1/N$, representing a Heisenberg limit to thermometry.
\end{abstract}

\maketitle

Temperature is a surprisingly subtle concept: it is easily percieved, yet it took 300 years to formalise the concept in operationally useful terms \cite{chang2004inventing}.  One of the principal challenges in understanding temperature and thermodynamics was the development of practical and accurate thermometers capable of measuring thermodynamic temperature.  Indeed, this remains an outstanding issue in modern thermometry: the relative uncertainty  in the CODATA value for $k_B$ is $u_r=1.7\times 10^{-6}$ \cite{RevModPhys.80.633}.  This makes it one of the least precisely measured  fundamental constants, surpassed only by Newton's Gravitational constant for imprecision in the CODATA values.

With a few exceptions, thermometers generally consist of a system of low thermal mass which is brought into thermal equilibrium with a bath of interest.  Following thermalisation of the thermometer, some state parameter of the thermometer is measured from which the temperature of the thermometer, and thus the bath, is inferred.  A low thermal mass is required to minimise the heat transferred between the bath and the thermometer compared to the total internal energy of the bath, so that the temperature of the bath is only slightly perturbed.  

For thermometric standards, the requirement of small thermal mass is less important.  The current CODATA value of $k_B$ is derived from measurements of a device embedded in an ice/water bath held at the triple point of water (TPW), which is defined to be exactly \mbox{$\tpw=273.16$ K} \cite{PhysRevLett.60.249}.  At the triple point, the temperature of the bath is independent of the internal energy of the bath, so the relative sizes of the thermometer and the bath are largely irrelevant.  In practise however, having a thermometer of small spatial extent is important for technical reasons: minimising temperature and pressure variations is easier over a smaller volume.  

Given the inherent advantages of small thermometers to both thermometry and thermometric standards, it is natural to enquire about the ultimate constraints that thermometer size imposes on the precision in measuring $\beta^{-1}=k_B T$. 
 In this letter, I first establish that thermalising thermometers are intrinsically shot-noise limited, i.e.\ the precision of such a thermometer comprising $N$ particles cannot scale better than $\sigma_\beta\sim1/\sqrt{N}$, assuming only that the total internal energy is an extensive parameter.  This shot-noise scaling is demonstrated in a very simple model consisting of $N$ independent atoms.  Extending this simple model, I introduce a new class of thermometer that  does not attain thermal equilibrium with the bath it is measuring.  Instead, the temperature of the bath is mapped onto a phase factor in the quantum state of the thermometer, which is then estimated using  interferometric methods.  By exploiting techniques developed for optimal phase estimation, I show that this non-thermalising thermometer is capable of attaining Heisenberg limited scaling, i.e. $\sigma_\beta\sim1/{N}$.

To establish the shot noise limit in thermalising thermometers, consider a thermometer in the thermal state
\begin{equation}
\rho=e^{-\beta H}/Z,
\end{equation}
where $Z=\tr\{{e^{-\beta H}}\}$.  Assuming that the average internal energy of the thermometer is extensive, it is given by $\bar E=-\frac{\partial \ln Z}{\partial \beta}=N\bar\varepsilon(\beta)$, where $\bar\varepsilon$ is the average internal energy per particle, which is independent of $N$. 
Since $\bar \varepsilon(\beta)$ is a monotonically increasing function of temperature, the sample standard deviation of $\bar \varepsilon$ and $\beta$ are related by the identity
\begin{equation}
\sigma_\beta ={\sigma_\varepsilon}/{\bar \varepsilon'},\label{eqnbeta}
\end{equation}
where $\bar \varepsilon'=|d\bar \varepsilon/d\beta|$. 
 To calculate ${\sigma_\varepsilon}$, note that the sample variance in the total internal energy is given by 
\begin{equation}
\sigma_E^2=\frac{\partial^2 \ln Z}{\partial \beta^2}=-\frac{\partial \bar E}{\partial \beta}=N\bar \varepsilon'.
\end{equation}
which demonstrates that the variance in the total internal energy of the thermometer is extensive.  The relative uncertainty in the energy per particle is thus 
\begin{equation}
\frac{\sigma_\varepsilon}{\bar \varepsilon}= \frac{\sigma_{E}}{\bar E}=\frac{1}{\sqrt{N}}\frac{\sqrt{\bar\varepsilon'}}{\bar \varepsilon}\label{eqnsepsilon}
\end{equation}
Equations (\ref{eqnbeta}) and (\ref{eqnsepsilon}) together imply that
\begin{equation}
{\sigma_\beta}{}=\frac{1}{\sqrt{N}}\frac{1}{\sqrt{\bar\varepsilon'}},\label{eqn:relunc}
\end{equation}
thus establishing the shot noise limit on extensive, thermalising thermometers.

This result may also be derived without explicit reference to any measurement parameter by computing the Fisher information, $F(\beta)$, for a thermal state. Since the thermal state is diagonal in the energy eigenbasis, it is straightforward to show that $F(\beta)=\sigma_E^2$.  It is then apparent that \eqn{eqn:relunc} saturates the Fisher inequality 
$
\sigma_\beta\geq1/{F(\beta)}
$  \cite{wiseman2009quantum,kok2010introduction}, and so \eqn{eqn:relunc} indeed represents the ultimate limit of the precision of a thermalising thermometer.

This scaling is potentially a significant issue for high precision thermometry. One promising avenue currently being pursued to improve the precision of measurements of $k_B$ 
relies on Doppler gas thermometry, in which high precision spectroscopy of a gas in thermal equilibrium with a TPW bath reveals the Maxwell-Boltzmann distribution of velocities \cite{PhysRevA.81.033848,PhysRevLett.98.250801,PhysRevLett.100.200801}, whose width is a direct measure of $k_B T_\mathrm{TPW}$. 
In recent Doppler thermometry experiments in alkali vapours the atomic flux through a beam of \mbox{$\sim10$} cm length and \mbox{$~2$ mm} diameter is $\dot N\sim 10^{15}$ atoms/sec \cite{Gar-Wing-Truong2010aa}.  The limit to the precision of such a thermometer is then $\sigma_\beta\sim({\dot N\tau})^{-1/2}\approx10^{-7.5}\tau^{1/2}$, where $\tau$ is the integration time.  At 1 second this sets a maximum precision in measurements of $k_B$ of 1 part in $10^{7.5}$, which is about 1.5 orders of magnitude better than the current CODATA estimates for $k_B$.
 
\begin{figure}
\begin{center}
\includegraphics[width=\columnwidth]{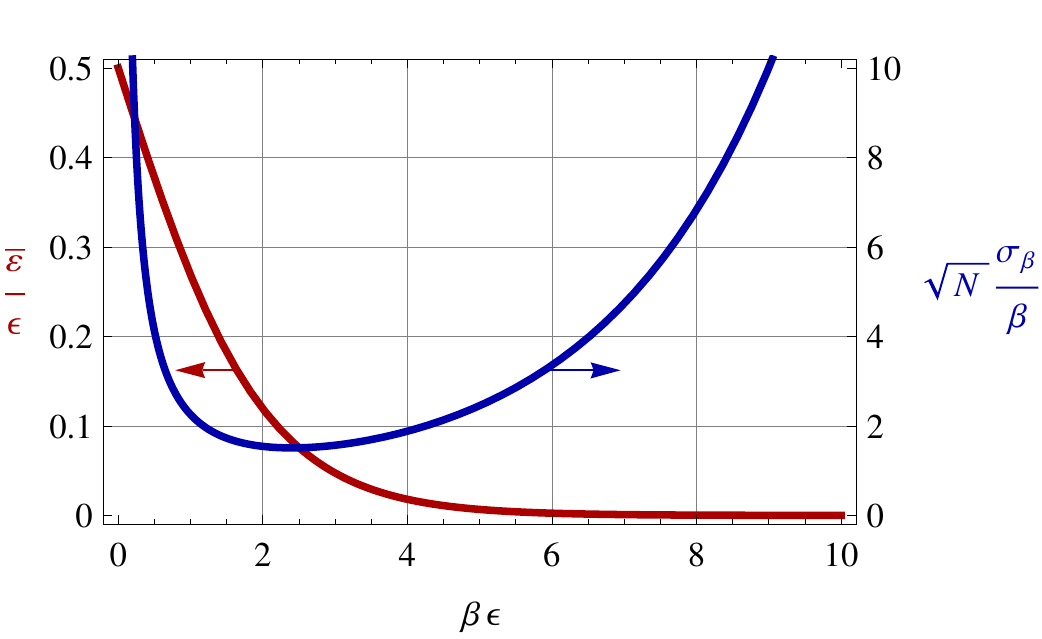}
\caption{Energy per particle, $\bar \varepsilon$, and relative sample uncertainty in $\beta$, scaled by $\sqrt{N}$, given by \eqn{eqn:relunc}, for the toy thermometer consisting of $N$ independent, two-level systems.} \label{fig1}
\end{center}
\end{figure}

The shot noise scaling of a thermalising thermometer is exhibited by a toy model of $N$ identical, non-interacting, two-level atoms, each with energy eigenstates $\ket{g}$ and $\ket {\epsilon}$.  The Hamiltonian is simply
\begin{equation}
H_{th}=\sum_{j=1}^N H_j,\textrm{ where }H_j=\epsilon\ket{\epsilon}_j.
\end{equation}
To simplify the analysis, suppose the thermometer atoms first come into thermal equilibrium with the bath, then they are adiabatically isolated from the bath, and finally the total energy of the atoms is measured to infer the temperature.  The partition function for the thermometer atoms is
\begin{equation}
Z=\tr \{e^{-\beta H_{th}}\}=(1+e^{-\beta \epsilon})^N
\end{equation}
and the internal energy and its variance are
\begin{eqnarray}
\bar E&=&\langle H_{th} \rangle=-\frac{\partial \ln Z}{\partial \beta}=N\frac{\epsilon}{1+e^{\beta \epsilon}},\\
\sigma_E^2&=&\langle H_{th}^2 \rangle-\langle H_{th} \rangle^2=\frac{\partial^2\ln Z}{\partial \beta^2}=N\frac{\epsilon^2e^{\beta\epsilon}}{(1+e^{\beta \epsilon})^2},
\end{eqnarray}
which are explicitly extensive.  By inspection, $\bar\varepsilon=\frac{\epsilon}{1+e^{\beta \epsilon}}$, and this, along with the relative uncertainty in $\beta$, are plotted in \fig{fig1}.  

This scaling behaviour,  $\sigma_\beta\propto 1/\sqrt{N}$, is strongly reminiscent of the standard quantum limit to phase estimation of an unknown phase, $\phi$, in one branch of an interferometer.  The standard quantum limit is a consequence of shot noise intrinsic  to  the states incident on the input ports of the interferometer.  In particular, if one input port of the interferometer is illuminated with a Fock state $\ket{n}$ of $n$ photons, or with a coherent pulse  $\ket{\alpha}$  with an expected number of photons $\alpha^2=n$, $\phi$ may be estimated to a precision that scales as $\sim1/\sqrt{n}$. 

It is possible to beat the standard quantum limit scaling for phase estimation by generating either squeezed states or entangled NOON states in the interferometer \cite{PhysRevD.23.1693,kok2010introduction}.  Indeed,  these non-classical states saturate the ultimate Heisenberg limit for phase estimation, with a precision that scales as $1/n$.

Given the superficial similarity between the $1/\sqrt{n}$ scaling of the precision of phase estimation in the standard quantum limit  and the $1/\sqrt{N}$ scaling of the precision of thermalising thermometers, it is natural to ask whether it is possible to construct a ``Heisenberg limited thermometer", whose precision improves  as $1/N$, just as the precision of Heisenberg limited phase estimation improves as $1/n$.  In what follows,  this question is answered in the affirmative.

It is clear from the preceding  analysis that \emph{any} extensive, thermalising thermometer will be subject to the scaling established above. In order to beat this scaling it is therefore necessary to relax either extensivity (of the internal energy) or thermalisation (of the thermometer).  The former seems difficult, since it requires interactions within the thermometer to dominate at all scales.  On the other hand, the construction presented below demonstrates that it is possible to do thermometry without thermalising the thermometer\footnote{Existing pyrometers measure temperatures that far exceed the melting point of the device components.}, and that such thermometers may indeed attain the Heisenberg limit $1/N$.  This  is accomplished by mapping the problem of measuring the temperature of a bath to the problem of estimating an unknown phase, then using techniques for optimal phase estimation. 

To demonstrate the function of a non-thermalising  thermometer,  a toy model for the thermometer, the bath and their interaction is now presented.  The thermometer is as described above: a collection of $N$ non-interacting, two-level atoms.  The bath is similar, consisting of \mbox{$M\gg N$} such atoms.  The form of interaction between the thermometer and the bath is crucial:
\begin{equation}
H_{int}=\alpha\sum_{j=1}^N\sum_{k=1}^M \ket{\epsilon}_j\bra{\epsilon}\otimes\ket{\epsilon}_k\bra{\epsilon},\label{int}
\end{equation}
where index $j$ refers to atoms in the thermometer and $k$ refers to atoms in the bath. 

The thermometer atoms are assumed to be bosonic, and creation and annihilation operators $a^\dagger_i$ and $a_i$ are introduced for atoms in mode $i$ acting on the atomic vacuum state $\ket{0}$.

\begin{figure}
\begin{center}
\includegraphics[width=\columnwidth]{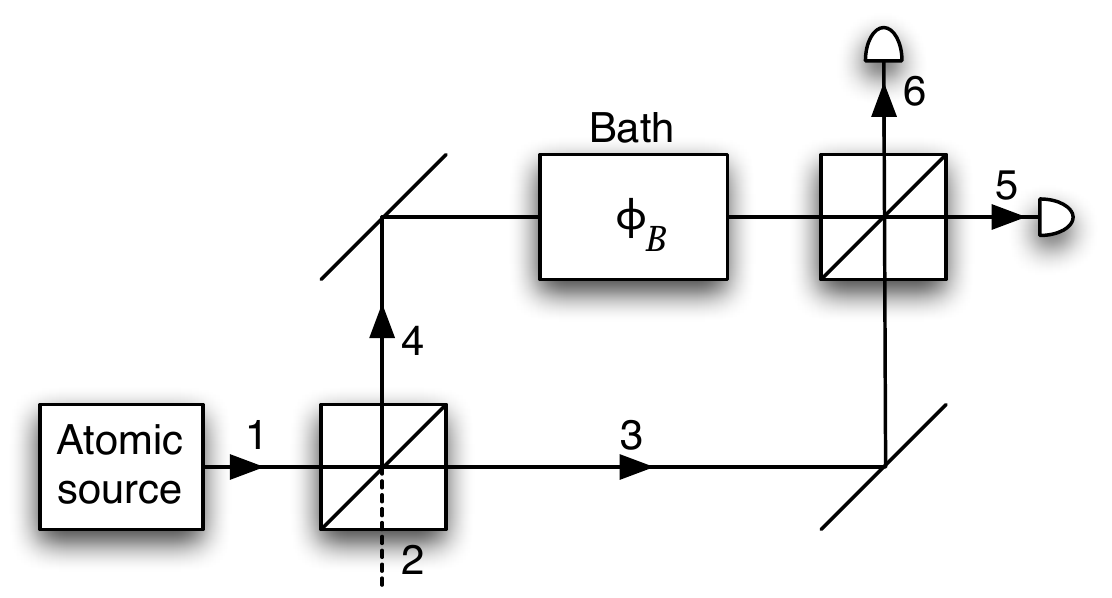}
\caption{An atomic interferometer with the bath in one branch.  Atoms from the thermometer are input into mode 1 and detected in modes 5 and 6.  Mode 2 is the vacuum port.} \label{fig2}
\end{center}
\end{figure}

The bath is assumed to be in a thermal state at some unknown temperature, $\rho_{B}=e^{-\beta H_{bath}}/Z_{bath}$, isolated from any environment, so that its internal energy is constant during the thermometry, and  embedded in an interferometer, as depicted in \fig{fig2}.   Suppose a single thermometer atom is prepared in the state $\ket{\epsilon}$ and introduced to the input port, 1, of the atomic beam splitter which effects the map
\begin{equation}
a^\dagger_1\rightarrow (a^\dagger_3+i\,a^\dagger_4)/\sqrt{2}.
\end{equation}
  As the wavefunction propagates through the interferometer the component in the branch containing the bath acquires a phase $\phi_B=\alpha \,m\,\tau$, where $m$ is the the number of excited atoms in the bath and $\tau$ is the interaction time.  

By repeating this $N$ times, $\phi_B$ and $m=\phi_B/(\alpha \tau)$ can be inferred from the output statistics.  This yields an estimate of both $\phi_B$ and $m$ with a precision given by 
\begin{equation}
\sigma_m^{(SN)}=\frac{\sigma_{\phi_B}^{(SN)}}{\alpha\tau}=\frac{1} {\alpha\tau}\frac{1}{\sqrt{N}}.
\end{equation} 
Since $\langle m\rangle=M/(1+e^{\beta \epsilon})$, the precision in the measured temperature is shot noise limited: 
%
\begin{eqnarray}
\sigma_\beta^{(SN)}&=&\sigma_m^{(SN)}/|d\langle m\rangle/d\beta|\nonumber\\
&=&\frac{1+e^{\beta\epsilon}}{\epsilon \,e^{\beta\epsilon}}\frac{1}{\alpha\tau}\frac{1}{\sqrt{N}}
\end{eqnarray}

Whilst this scaling is no better than that of a thermalising thermometer,  this construction  demonstrates that measuring the bath temperature using an $N$-atom thermometer can be mapped onto the task of phase estimation using $N$ atoms.  It follows that techniques from optimal phase estimation lead to  improved precision in thermometry.

In particular, suppose the linear atomic beam splitters shown in \fig{fig2} are replaced with non-linear beam splitters that effect the  $N$-atom map
\begin{equation}
(a^\dagger_1)^N\ket{0}\rightarrow  ((a^\dagger_3)^N+i(a^\dagger_4)^N)\ket{0},
\end{equation}
which maps an $N$ photon fock state onto a NOON state in which all $N$ atoms take the same branch.  As the atoms in the upper branch interact with the bath atoms, according to the toy Hamiltonian given above, a phase $N\phi_B$ is accumulated on mode 4:
\begin{equation}
\rightarrow  ((a^\dagger_3)^N+i\,e^{iN\,\phi_B}(a^\dagger_4)^N)\ket{0},
\end{equation}
As shown in \cite{PhysRevA.65.052104},  this state has maximal Fisher information with respect to $\phi_B$, so $\phi_B$ can be measured at the Heisenberg limit, i.e.\ with precision $\sigma_{\phi_B}^{(H)}=1/N$.  This is accomplished by measuring the observable 
\begin{equation}
\hat A_N=(a^\dagger_3)^N\ket{0}\bra{0}(a_3)^N+(a^\dagger_4)^N\ket{0}\bra{0}(a_4)^N.
\end{equation}
It follows that the temperature can be measured with a precision
\begin{eqnarray}
\sigma_\beta^{(H)}&=&\sigma_m^{(H)}/|d\langle m\rangle/d\beta|\nonumber\\
&=&\frac{1+e^{\beta\epsilon}}{\epsilon \,e^{\beta\epsilon}}\frac{1}{\alpha\tau}\frac{1}{N}
\end{eqnarray}

This demonstrates an $N$ atom thermometer that can indeed attain the Heisenberg limit, beating the shot noise limit.  In this construction, there were two crucial assumptions.  Firstly, the bath was assumed to be isolated, so that both $m$ and $M$ remain fixed during the measurement.  If $m$ varies during the measurement, then fluctuations in $N\phi_B$ would lead to dephasing of the NOON state, and the precision would be reduced.  If $M$ varies, then clearly $m$ will also, but so does the estimate of the total internal energy of the system.

Secondly, the interaction between the atoms in the bath and those in the thermometer must be precisely engineered, to be of the form given in \eqn{int}, which is diagonal in the energy eigenbasis.  This means that the thermometer is not universal, since it must be designed to be compatible with the bath.  This is in contrast to thermalising thermometers, which attain thermal equilibrium for a very large class of bath systems.  Indeed, it suggests that for thermometric standards, in which the bath is chosen for convenient properties of its triple point, it is a problem of extreme complexity to engineer a coupling of the form required.
Whilst there may be alternative models of thermometry that also attain the Heisenberg limit, it is not clear that these two assumption can be relaxed. 

As discussed above, the temperature of an isolated thermal bath is intrinsically defined only to a precision $\delta_{\beta_B}\sim1/\sqrt{M}$.  It follows that the maximum precision required of a thermometer is limited by the size of the bath it is measuring: no advantage is gained by using a thermometer capable of higher precision, nor in extending the integration time for improved averaging.  The fact that thermalising thermometers are shot noise limited, $\delta^{(SN)}_{\beta_T}\sim1/\sqrt{N}$, implies that to reach the intrinsic precision determined by $\delta_{\beta_B}$ such thermometers need to be at least as large as the bath they are measuring, i.e. 
\begin{equation}
\delta^{(SN)}_{\beta_T}=\delta_{\beta_B}\Rightarrow N\sim M.
\end{equation}
This condition competes with the requirement that the thermometer be much smaller than the bath, highlighting the implicit tension between high-precision thermometry and thermometer-induced perturbations of the bath.

In contrast, for a Heisenberg limited thermometer, for which $\delta^{(H)}_{\beta_T}\sim1/{N}$, the condition
\begin{equation}
\delta^{(H)}_{\beta_T}=\delta_{\beta_B}\Rightarrow N^2\sim M.
\end{equation}
is satisfied for $N\ll M$, and so  the thermometer may be much smaller than the bath.  

Small scale demonstrations of Heisenberg limited thermometry using the ideas discussed here may be realised in atom or ion traps, e.g.\ in a trap consisting of a relatively few bath atoms (say $M\lesssim100$), and even fewer thermometer atoms ($N\approx\sqrt{M}\lesssim 10$), possibly leading to new techniques to measure the extremely low temperatures achieved in these systems.

Having established, by explicit construction, the existence of Heisenberg limited thermometers, there are several natural question that follow from this work.  It seems plausible that the Heisenberg limit is the ultimate limit of thermometry, however the constructive arguments given here do not establish this definitively.  Further, the constraints on the bath and the interaction Hamiltonian in the toy model required  to attain the Heisenberg limit of thermometry are potentially difficult to engineer.  There may be other implementations that relax these constraints leading to more practical Heisenberg limited thermometers.  Finally there is an intriguing connection with thermal clocks \cite{Milburn2010aa} that remains to be explored.

This work was funded by the Australian Research Council. I thank Gerard Milburn, Andre Luiten, Gar-Wing Truong, Eric May and Geoff Pryde  for helpful conversations.

\bibliography{QuantumThermometry}


\end{document}